\documentclass[traditabstract]{aa} 
\usepackage{epsfig,color}
\usepackage{amsmath}
\usepackage{mathptmx}
\usepackage{amssymb}
\usepackage{units}
\usepackage{natbib}
\usepackage[varg]{txfonts}
\usepackage[tight]{subfigure}

\newcommand{\LEAFS}{{\sc{Leafs}}}
\newcommand{\ARTIS}{{\sc{Artis}}}
\newcommand{\gcc}{\,\mathrm{g\ cm^{-3}}}

\newcommand{\kms}{\,\mathrm{km\ s^{-1}}}
\newcommand{\cm}{\,\mathrm{cm}}
\newcommand{\km}{\,\mathrm{km}}

\newcommand{\nuc}[2]{\ensuremath{\mathrm{^{#1}#2}}}
\newcommand{\ions}[2]{#1\,{\sc #2}}
\newcommand{\ye}{\ensuremath{Y_\mathrm{e}}}
\newcommand{\msun}{\ensuremath{\mathrm{M}_\odot}}
\newcommand{\mch}{\ensuremath{\mathrm{M_{Ch}}}}
\begin{document}
\title{Three-dimensional simulations of gravitationally confined
  detonations compared to observations of SN~1991T}
\author{Ivo~R.~Seitenzahl$^{1,2,3}$,
 Markus~Kromer$^{4}$, 
 Sebastian~T.~Ohlmann$^{5}$, 
 Franco~Ciaraldi-Schoolmann$^{1}$,
 Kai~Marquardt$^{5,6}$,
 Michael~Fink$^{6}$, 
 Wolfgang~Hillebrandt$^{1}$,
 R\"udiger~Pakmor$^{5}$,  
 Friedrich~K.~R\"opke$^{5,7}$, 
 Ashley~J.~Ruiter$^{2,3}$,        
 Stuart~A.~Sim$^{8}$, 
 Stefan Taubenberger$^{1}$}

\institute{
Max-Planck-Institut f\"ur Astrophysik,
Karl-Schwarzschild-Stra{\ss}e 1, 85748 Garching, Germany\\
\email{ivo.seitenzahl@anu.edu.au}\and
Research School of Astronomy and Astrophysics, The Australian National University, Canberra, ACT 2611,
Australia\and
ARC Centre of Excellence for All-Sky Astrophysics (CAASTRO)\and
The Oskar Klein Centre \& Department of Astronomy,
       Stockholm University, AlbaNova, SE-106 91 Stockholm, Sweden\and
Heidelberger Institut f\"{u}r Theoretische Studien,
  Schloss-Wolfsbrunnenweg 35, 69118 Heidelberg, Germany\and
Institut f\"ur Theoretische Physik und Astrophysik, Universit\"at
   W\"urzburg, Emil-Fischer-Stra{\ss}e 31, 97074 W\"urzburg\and
Zentrum f{\"u}r Astronomie der Universit{\"a}t Heidelberg, 
       Institut f{\"u}r Theoretische Astrophysik, Philosophenweg 12, 
       D-69120 Heidelberg, Germany\and
Astrophysics Research Centre, School of Mathematics and Physics, 
       Queen's University Belfast, Belfast BT7 1NN, UK
}

\date{Received xxxx xx, xxxx / accepted xxxx xx, xxxx}


\abstract{The gravitationally confined detonation (GCD) model has been
  proposed as a possible explosion mechanism for Type Ia supernovae in
  the single-degenerate evolution channel. It starts with
  ignition of a deflagration in a single, off-center bubble in a
  near-Chandrasekhar-mass white dwarf.  Driven by buoyancy,
  the deflagration flame rises in a narrow cone towards
  the surface. For the most part, the main component of the flow of the expanding ashes remains radial, but upon reaching the outer, low-pressure layers of the white dwarf, an additional lateral component develops. This makes the deflagration ashes converge again at the opposite side, where the compression heats fuel and a detonation may be launched.
  We first perform five three-dimensional
  hydrodynamic simulations of the deflagration phase in $1.4\,\msun$
  carbon/oxygen white dwarfs at intermediate-resolution ($256^3$ computational zones).
  We confirm that the closer the initial deflagration is ignited to the center, the slower the buoyant rise and the
  longer the deflagration ashes takes to break out and close in on the opposite pole to collide.
  To test the GCD explosion model, we then perform a high resolution ($512^3$ computational zones) simulation for a model
  with an ignition spot offset near the upper limit of what is still justifiable, $200\,\km$.
  This high-resolution simulation meets our deliberately optimistic detonation criteria and we initiate a detonation.
  The detonation burns through the white dwarf and leads to its complete
  disruption. For this model, we determine detailed nucleosynthetic
  yields by post-processing $10^6$ tracer particles with a 384 nuclide
  reaction network and we present multi-band light curves and
  time-dependent optical spectra.  We find that our synthetic observables
  show a prominent viewing-angle sensitivity in ultraviolet and
  blue wavelength bands, which is in tension with observed SNe~Ia.
  The strong dependence on viewing-angle is caused by the asymmetric
  distribution of the deflagration ashes in the outer ejecta layers.
  Finally, we perform a comparison of our model to SN~1991T.
  The overall flux-level of the model is slightly too low and the model predicts
  pre-maximum light spectral features due to Ca, S, and Si that are too strong.
  Furthermore, the model chemical abundance stratification
  qualitatively disagrees with recent abundance tomography results in two key
  areas: our model lacks low velocity stable Fe and instead has copious amounts
  of high-velocity \nuc{56}{Ni} and stable Fe. We therefore do not find good agreement
  of the model with SN~1991T.}


\titlerunning{3D GCD model vs. SN~1991T}
\authorrunning{Seitenzahl et al. 2015}
\maketitle

\section{Introduction}
\label{sec:int}
It is widely accepted that Type Ia supernovae (SNe~Ia) are
thermonuclear explosions of white dwarf stars. Since isolated white
dwarfs are stable objects, the progenitor star must be
interacting with a companion such that critical conditions
necessary for explosion can be achieved. These critical conditions vary depending on the explosion
mechanism, of which several have been proposed.

The manner in which the exploding, probably carbon-oxygen (CO), white
dwarf (WD) accretes matter in the first place and the nature of the
companion have been topics of debate for decades
\citep{whelan1973a,nomoto1982a,iben1984a,webbink1984a,yungelson1995a,tutukov1996a,yungelson2000a,han2004a,ruiter2009a,mennekens2010a,toonen2012a,ruiter2013a,hillebrandt2013a}.

Different progenitor evolution scenarios exist, and for some scenarios yet again different
explosion models have been proposed. For the single-degenerate
scenario, e.g., pure deflagrations, deflagration-to-detonation
transitions, pulsational reverse detonations, and gravitationally
confined detonation (GCD) models have been suggested.
Here we focus on the GCD model, first discussed by \citet{plewa2004a}.
The GCD model starts with the ignition of a deflagration in a single, off-center 
bubble in a near-Chandrasekhar-mass (near-\mch) white dwarf (WD).  Driven by buoyancy, the burning products of the deflagration rise quickly towards 
the stellar surface, where, in addition to the dominant radial expansion, they develop a lateral velocity component and converge at the opposite side.
There, the flow compresses and heats still unburned fuel and a
detonation may be launched. For a successful triggering of the
  detonation, flow convergence in rather dense material is
  required. Therefore, a weak deflagration and only a modest expansion
  of the WD is expected to favor the scenario (see Section~\ref{sec:hydro3}). Consequently, the ensuing detonation burns a large amount of fuel at high densities, mostly to \nuc{56}{Ni}, thus producing bright SNe.

SN~1991T \citep{filippenko1992a,phillips1992a,schmidt1994b,lira1998a} is the prototypical event of a spectroscopically peculiar class of energetic and luminous SNe~Ia. SN 1991T is the best characterized exemplar of a sub-class (SNe~91T) of SNe~Ia making up a few percent of all observed SNe~Ia \citep{li2011a,silverman2012b,blondin2012a}. SNe~91T also are known to occur preferentially in late-type galaxies \citep{hamuy2000a,howell2001b}, indicating an origin in young stellar populations. SNe~91T further clearly distinguish themselves from normal SNe~Ia by their peculiar pre-maximum light spectra. In particular, the characteristic \ions{Si}{ii} $\lambda\lambda 5972, 6355$ features prominent in normal SNe~Ia \citep[e.g.,][]{branch1993a} are essentially absent before maximum light; the same holds for other features of intermediate mass elements, such as \ions{Ca}{ii} or \ions{S}{ii}. Instead, early spectra show unusual \ions{Fe}{iii} features \citep{filippenko1992a,ruiz-lapuente1992a, jeffery1992a}, requiring high ionization and Fe abundance at high velocity. Moreover, SNe 91T events are more luminous by about 0.2--0.3 mag than the width-luminosity relation predicts for normal SNe Ia \citep{blondin2012a}.

At first sight, it is tempting to liken the GCD models with SNe~91T \citep[e.g.,][]{fisher2015a}. The deflagration ashes present at high
velocities in the outer layers of GCD models results in a
chemically mixed composition highly enriched with Fe-group isotopes at
high velocities, which could explain key results obtained from
interpreting observations of SN~1991T
(e.g., \citealt{mazzali1995a,fisher1999a}; but see \citealt{sasdelli2014a}).
Furthermore, the fact that only a few per cent
of SNe~Ia are classified as 1991T-like \citep{li2011a,silverman2012b,blondin2012a} would be naturally
explained by the relative scarcity of the single-degenerate scenario
compared to, e.g., the violent-merger or double-detonation scenarios
\citep[e.g.,][]{ruiter2009a,ruiter2011a,ruiter2013a,fisher2015a}.
The identification of GCD explosions with SNe~91T is facing a challenge
from observations of the supernova remnant SNR~0509-67.5 in the LMC.
First, based on light echo spectra, \citet{rest2008b} demonstrated clearly that this SN was a 1991T-like explosion.
Second, \citet{schaefer2012a} rule out the existence of a companion star in this SNR to very deep limits and thereby argue for a double-degenerate progenitor, which would exclude the canonical single-degenerate formation channel as a path to the GCD model. Note that recently \citet{garcia2016a} have questioned the viability of the mechanism altogether, arguing that the Coriolis-force substantially breaks the symmetry and thereby disfavours the emergence of a detonation. 

Historically, both delayed-detonations in near-\mch\ WDs
\citep{mazzali1995a} and sub-Chandrasekhar-mass (sub-\mch) double detonations (DDs)
\citep[e.g.,][]{liu1997a} have been suggested as possible explosion
models for SN~1991T.
However, detailed comparisons of SN~1991T and
synthetic spectra and light curves of bright sub-\mch\ DD models \citep{kromer2010a,woosley2011b}
and bright near-\mch\ delayed-detonation models \citep{sim2013a} 
found substantial disagreement between the model predictions and the observations.
Similarly, pure detonations of ONe WDs \citep{marquardt2015a}, which can produce SNe~Ia with \nuc{56}{Ni} masses around $1\,\msun$, are not good models for SNe~91T (for example, the strong Si\,{\sc ii} and Ca\,{\sc ii} absorption features present in the models before maximum light are almost absent in SNe~91T). 
   
In spite of the fact that successful explosions of near-\mch\
WDs in the GCD-framework have been obtained in
several hydrodynamical explosion simulations
\citep{plewa2007a,meakin2009a,jordan2012a}, only \citet{meakin2009a}
present detailed isotopic yields for their 2D-models.
\citet{kasen2007b} presented broadband optical and near-infrared light
curves, spectral time series and spectropolarization for their Y12 model
and find a general success in reproducing the basic properties
of observed SNe~Ia.

In this work, we present (multi-band) light curves and time dependent
optical spectra for an explosion model representative of the class of
GCD models, for which we employed a detailed treatment of the
nucleosynthesis for the detonation as well as for the deflagration.
In Section~\ref{sec:hydro}, we briefly discuss our series of intermediate-resolution, single bubble,
off-center deflagration simulations and present the evolution of a
high-resolution model, which had met our very optimistic detonation criteria.
In Section~\ref{sec:obs}, we present spectra and light curves for this
model and compare them with observations. In Section~\ref{sec:summary}
we summarize and conclude that these synthetic observables do not
resemble any known subclass of SNe~Ia.

\section{Hydrodynamic explosion simulations}
\label{sec:hydro}
\begin{figure*}[ht!]
  \begin{center}
    \centering
    {\includegraphics{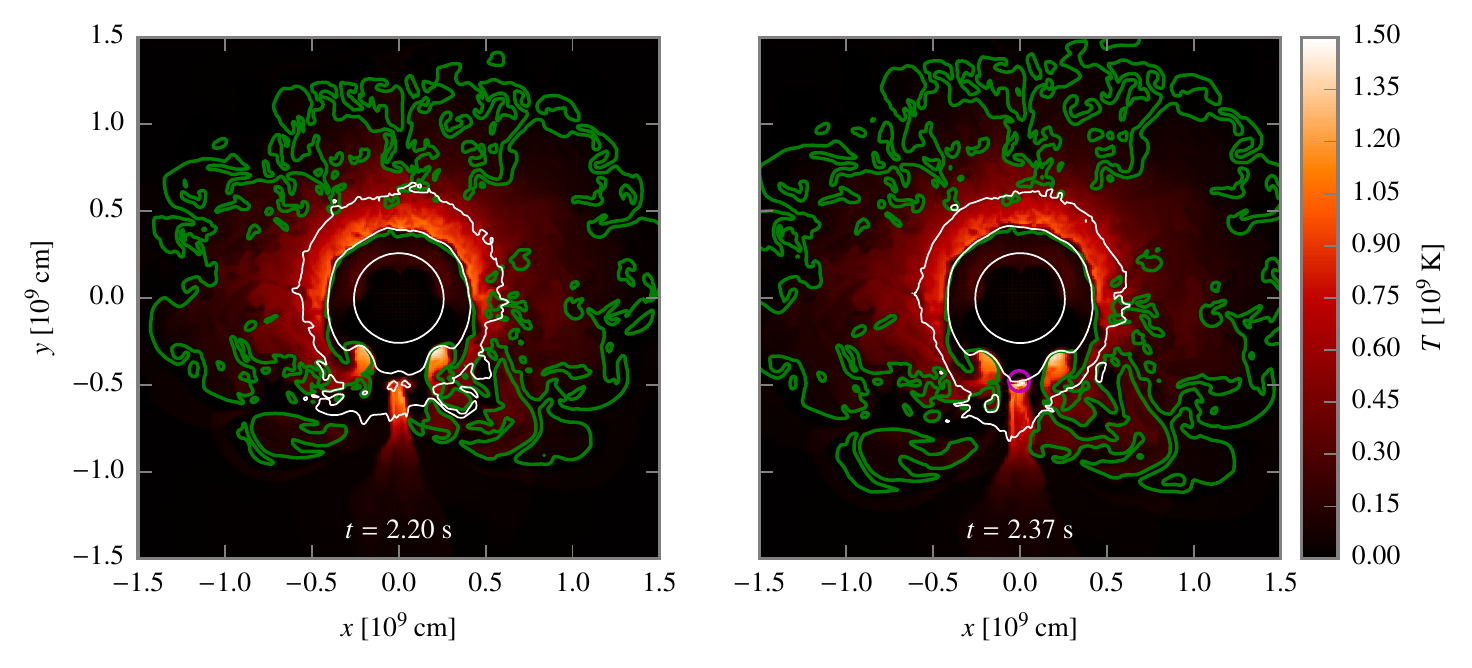}}
    \caption{2D slice through the midplane of the simulation volume along the
    $x$ and $z$-axes for model GCD200 at $t = 2.20$\,s (left panel) and at $t =
    2.37$\,s (right panel) with color coding temperature. The green contour line
    displays the position of the deflagration level set; anything enclosed by a green line has been burned in the deflagration.
    The white contour lines show the density at $10^7\gcc$, $10^6\gcc$, and $10^5\gcc$ (from the center
    outwards). The panel on the left shows the temperature slightly 
    before the detonation is initialized and the panel on the right shows the
    temperature at the time of the initialization of the detonation. The
    detonation spot is marked by a magenta circle (right panel).}
    \label{fig:temp}
  \end{center}
\end{figure*}
According to theory, a detonation may spontaneously be initiated if a
sufficiently shallow induction-time gradient can be set up to
facilitate the formation of a shock in the SWACER (shock wave
amplification through coherent energy release) mechanism \citep[see
e.g.,][]{seitenzahl2009b}.  In the GCD model, suitable induction-time
gradients may be obtained in the collision region near the stagnation
point of the surface flow, directly opposite to the point of breakout
of the rising deflagration bubble \citep{seitenzahl2009c}. Since the
spatial scales relevant to the initiation of a detonation cannot be
resolved in full star multi-dimensional explosion simulations
\citep[see the discussion in][]{seitenzahl2009c}, it is common
practice
\citep[e.g.,][]{jordan2008a,jordan2012a,guillochon2010a,pakmor2011b,pakmor2012a}
to pick a certain critical density $\rho_\mathrm{crit}$ and
temperature $T_\mathrm{crit}$ that a cell composed of nuclear fuel
must exceed for a detonation to be initiated.  However, these critical values
are no more than informed guesses based on separate
high-resolution, one-dimensional detonation initiation calculations
\citep[e.g.,][]{niemeyer1997a,seitenzahl2009b}.

\subsection{Computational method}
\label{sec:hydro1}
We perform three-dimensional full-star simulations with the
hydrodynamic supernova explosion code \LEAFS. The whole
exploding WD and the deflagration flame are captured separately on two
different nested co-moving grids \citep{roepke2005c, roepke2006a}.
Since the flame is very thin compared to the radius of a near-\mch\
WD, the former is treated as a sharp discontinuity
with the help of the level set technique \citep{osher1988a, reinecke1999a}. The
effective burning speed of the deflagration flame
is composed of a laminar and a turbulent contribution
\citep{pocheau1994a}.  For the former we use tabulated values from
\citet{timmes1992a}, while the turbulent flame speed is determined
from a subgrid-scale turbulence model of \citet{schmidt2006b,
  schmidt2006c}. The energy release behind the deflagration level set is
determined as a function of fuel density by interpolation in a table
that is consistent with the energy release of a detailed nuclear reaction network; for details see Appendix A of \citet{fink2014a}.
The detonation is also modeled with a (separate) level set. 
For the detonation speed and energy release we use the tables of
\citet{fink2010a}, which, just as the deflagration tables, take the effects of incomplete burning at low densities into account.
The reactive Euler equations are numerically solved
with a finite-volume method \citep{fryxell1989a} based on the piecewise-parabolic method (PPM) of
\citet{colella1984a}. For the nucleosynthesis, we post-process one
million tracer particles with a nuclear reaction network using the
technique described in \citet{travaglio2004a} and
\citet{seitenzahl2010a}.  The simulation code is essentially the same
as described in \citet{seitenzahl2013a}, a
key difference, however, is that the deflagration-to-detonation transition
module has been de-activated and the initiation of detonations is handled in
a different way.

Here, we initialize the detonation around the grid cell in which both
critical values $\rho_\mathrm{crit}$ and $T_\mathrm{crit}$ are
(simultaneously) reached or exceeded first.  Furthermore, to avoid
artificial numerical triggering of the detonation, we require that the
deflagration level set is at least one grid cell away and the grid
cell is composed of at least 90 per cent of unburned \nuc{16}{O} and \nuc{12}{C}.

\begin{figure*}
  \subfigure[GCD200, t = 2.37 s]
  {\includegraphics[height=0.31\textwidth,angle=-90]{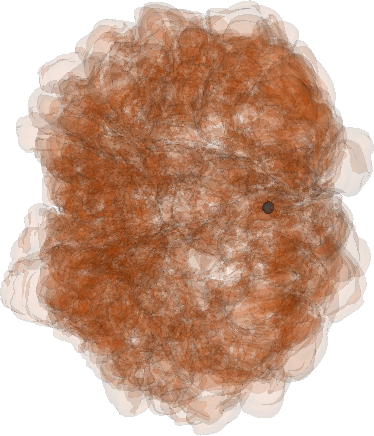}}
  \hfill
  \subfigure[GCD200, t = 2.4 s]
  {\includegraphics[height=0.31\textwidth,angle=-90]{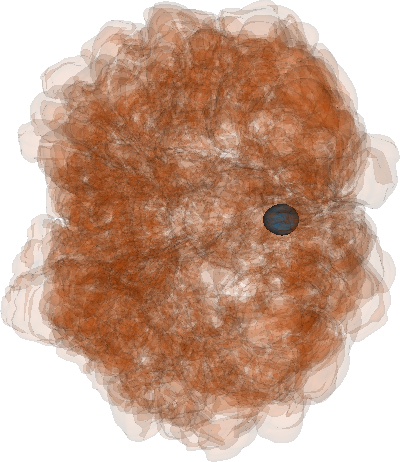}}
  \hfill
  \subfigure[GCD200, t = 2.5 s]
  {\includegraphics[height=0.31\textwidth,angle=-90]{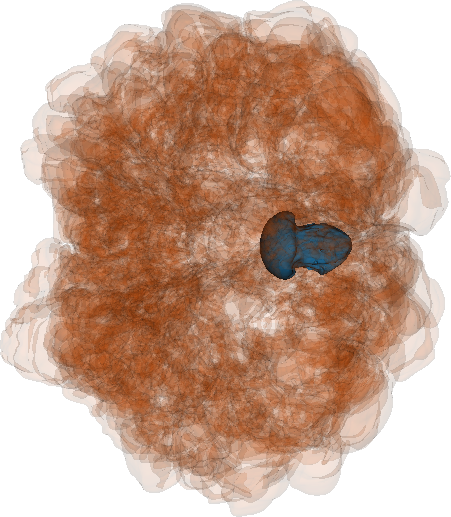}}
  \\
  \subfigure[GCD200, t = 2.7 s]
  {\includegraphics[height=0.31\textwidth,angle=-90]{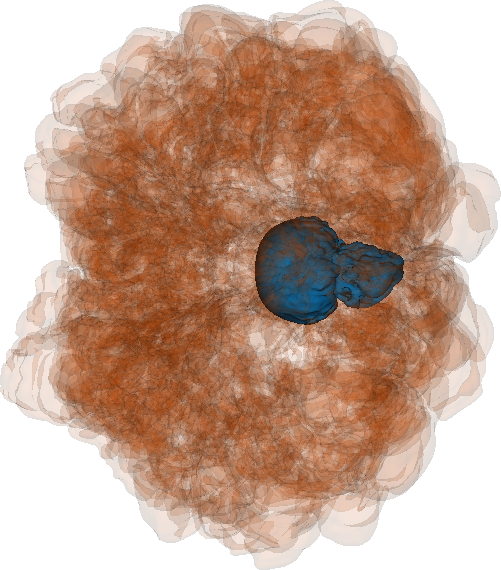}}
  \hfill
  \subfigure[GCD200, t = 3.9 s]
  {\includegraphics[height=0.31\textwidth,angle=-90]{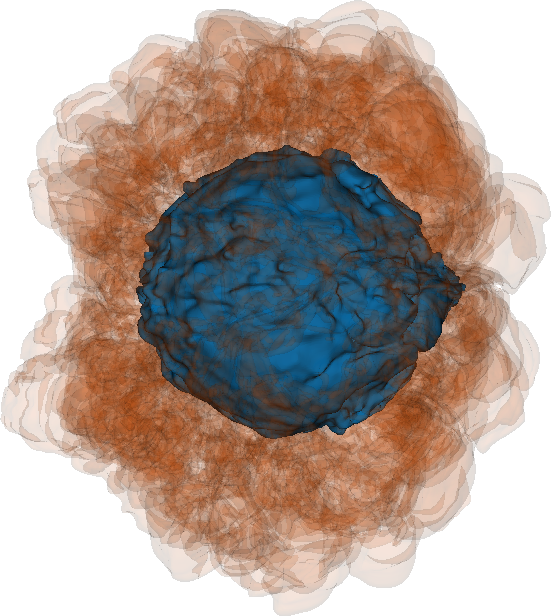}}
  \hfill
  \subfigure[GCD200, t = 60 s]
  {\includegraphics[height=0.31\textwidth,angle=-90]{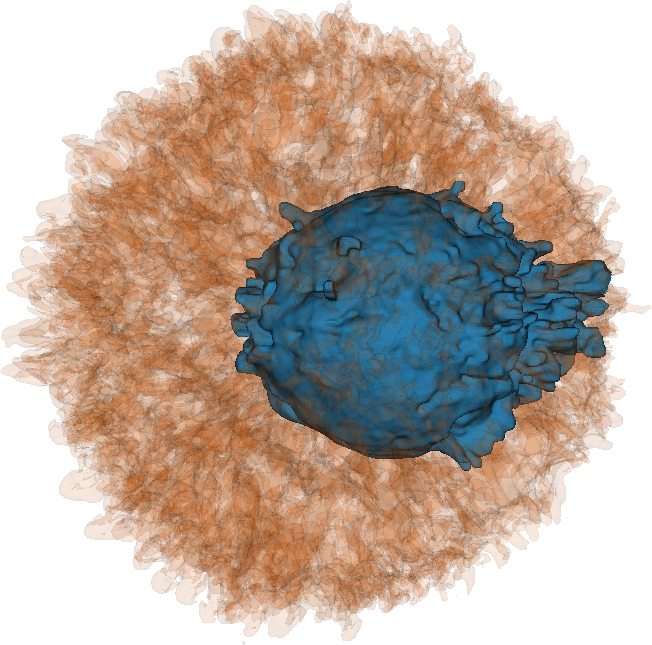}}
  \caption{Evolution of the GCD200 model for six different instances of time.
    The deflagration level set is shown in red and the detonation level set is
    shown in blue. Since the deflagration ashes almost completely surround the
    detonation, we have rendered the deflagration level set
    semi-transparent to allow for visualization of the enclosed detonation
    front. The spatial scales in the plots are $1.23\times 10^{10}$\ cm,
    $1.25\times 10^{10}$\ cm, $1.33\times 10^{10}$\ cm, $1.47\times 10^{10}$\ cm,
    $2.46\times 10^{10}$\ cm, and $1.59\times 10^{12}$\ cm for panels (a) to (f),
    respectively. 
  }
  \label{fig:gcd3}
\end{figure*}

\begin{figure*}
  \begin{center}
    \subfigure[]
        {\includegraphics[width=\columnwidth]{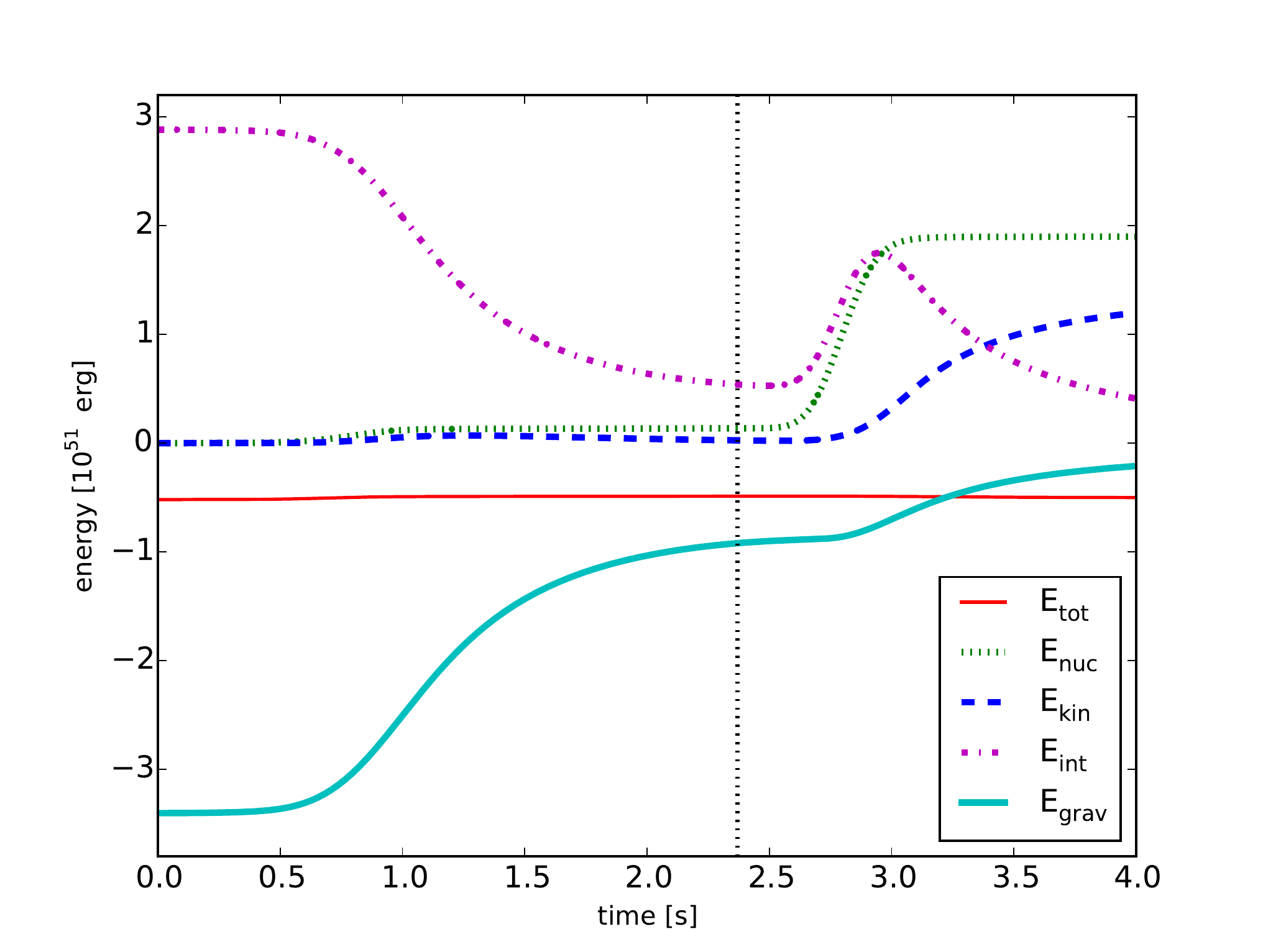}}
    \subfigure[]
        {\includegraphics[width=\columnwidth]{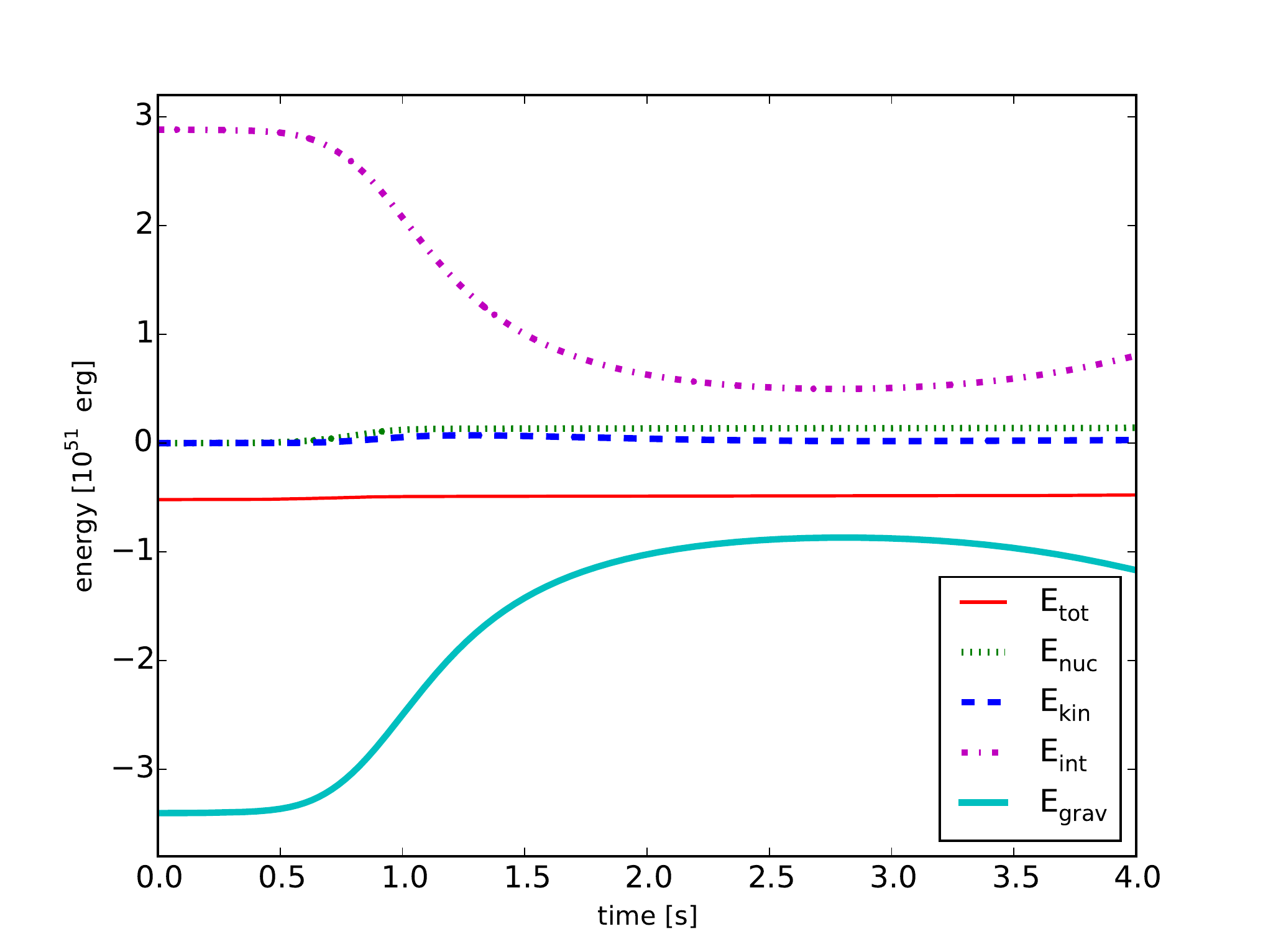}}
    \caption{Temporal evolution of $E_\mathrm{tot}$,
      $E_\mathrm{nuc}$, $E_\mathrm{kin}$, $E_\mathrm{int}$ and
      $E_\mathrm{grav}$. The left panel (a) shows model
      GCD200. The vertical line marks the time at which the
      detonation is initialized. The right panel (b) is for the
      corresponding pure deflagration case, if no detonation is
      initiated.}
    \label{fig:energies}
  \end{center}
\end{figure*}

Our initial stellar model is an isothermal near-\mch\ WD with a central density of $2.9\times 10^9\,\gcc$ and a
temperature of $5\times 10^5\,\mathrm{K}$. For the hydrodynamical
simulations, the stellar material is assumed to be composed of
carbon and oxygen where we approximate the effect of the assumed solar metallicity of the zero-age main-sequence progenitor on the stellar structure by initializing the electron fraction to $\ye = 0.49886$ \citep[see e.g.][]{seitenzahl2011a}.

\subsection{Moderate-resolution test cases}
\label{sec:hydro3}
We ignite the deflagration in a single spherical off-center
region with a bubble radius of $\unit[10^6]{cm}$ and we carry out
simulations of five models at a moderate resolution of $256^3$ grid
cells. The models vary only in the distance $d_\mathrm{k}$ of the
center of the ignition kernel to the center of the white dwarf and we have tried $d_\mathrm{k}=12,37,64,125, \mathrm{and}\, 200\,\km$.  
All ignitions occur at a density greater than $2.3\times 10^9\,\gcc$.

The closer the initial deflagration is ignited to the center, the slower the buoyant rise and the longer the deflagration ashes takes to break out and close in on the opposite pole to collide.
Since the white dwarf keeps expanding in the meantime, obtaining 
high temperatures in the collision region at densities higher than
$\rho_\mathrm{crit}$ becomes less probable.
We thus find, as in \citet{roepke2007a}, that larger distances of the ignition kernel from
the center of the white dwarf lead to more suitable conditions
for a detonation at the closing point of the deflagration ashes. The
reason is that the deflagration evolves faster, burns less mass and releases less energy,
which consequently leads to less expansion of the WD so that, when the erupted deflagration ashes converge,
sufficiently high temperatures are more readily obtained at densities where detonations can occur
\citep[][]{townsley2007a,roepke2007a,aspden2011a}.\footnote{See
  \citet{malone2014a} for an explanation of how background turbulence of the convective flow moderates this trend for very small offsets.}

\subsection{High-resolution simulation of model GCD200}
\label{sec:hydro4}
Our main motivation for this work is not to elucidate \emph{whether} a
detonation can be triggered in the GCD framework, but rather what the
predicted observables of the model look like \emph{if} a detonation is
triggered. We therefore chose a large offset for the ignition of the deflagration, $200\km$, which is more favorable to attaining detonation conditions than smaller offsets are. While $200\km$ is still commensurate with the analytical calculations of \citet{garcia1995a} and
\citet{woosley2004a}, the numerical simulations of \citet{kuhlen2006a} and \citet{nonaka2012a} favour much smaller offset distances. It is therefore questionable whether an offset of $200\km$ for the ignition point is actually attainable.

Past 3D hydrodynamics simulations with the \LEAFS\
code \citep[see table 4 of][]{roepke2007a} failed to obtain the more
conservative detonation conditions $\rho_\mathrm{crit} = 10^7 \gcc$
and $T_\mathrm{crit} = \unit[1.9\times 10^9]{K}$, even for offset distances
of the initial deflagration bubble as large as $200\,\km$. Since we focus here on the model implications \textit{if}
a detonation occurs, we choose lower, less restrictive values for this work: $\rho_\mathrm{crit} = 10^6 \gcc$ and $T_\mathrm{crit} = \unit[1.0\times 10^9]{K}$. We also require the fuel content of the cell to be greater than 90 per cent, i.e., $X_\mathrm{fuel} > 0.9$.
We note that this choice of critical detonation conditions is very optimistic, pushing the boundaries of
the parameter space where detonations are expected to form via the gradient mechanism \citep{seitenzahl2009b}.

We simulate a single bubble ignition model with bubble radius $10^6\cm$ and $d_\mathrm{k}=200\km$, with an increased resolution of $512^3$ grid cells (hereafter referred to as GCD200). This high-resolution model evolves to meet our critical detonation conditions (outlined above) at $t=2.37\,\mathrm{s}$ and we initialize a  single detonation around the grid cell where the constraints are first satisfied. The detonation is initiated by adding a second levelset at this instant with positive values in a sphere of a radius of three cells (330\,km) and converting this material instantaneously to iron group elements.

In Fig.~\ref{fig:temp}, we show a two-dimensional slice of the temperature
for two instants of time ($2.20\,\mathrm{s}$ and $2.37\,\mathrm{s}$).
Most material at densities above $10^6\,\gcc$ is unburned and the WD core (densities above $10^7\gcc$) appears spherical
and hardly distorted by the deflagration.
In the second snapshot, which corresponds to
the time when we initiate the detonation, a hotspot is forming in the
focus region in unburned material where the ashes of the deflagration converge.
The central density of the white dwarf at this time ($2.37\,\mathrm{s}$) is
$6.63\times 10^7\,\mathrm{g}\,\mathrm{cm}^{-3}$. 
The density and temperature in the
grid cell around which the detonation is initialized are
$\rho_\mathrm{fuel} = 1.09\times 10^6\, \mathrm{g}\,\mathrm{cm}^{-3}$ and
$T_\mathrm{fuel} = 1.02\times 10^{9}\,\mathrm{K}$.
These thermodynamic conditions only satisfy the critical detonation criteria for this work, but 
would not have satisfied the more restrictive critical conditions of \citet{roepke2007a}.
Although our model was far from the more canonical and arguably more realistic detonation criteria
$\rho_\mathrm{crit} \approx 1.0\times 10^7\, \mathrm{g}\,\mathrm{cm}^{-3}$ and
$T_\mathrm{crit} \approx 2.0\times 10^{9}\,\mathrm{K}$ \citep{seitenzahl2009b},
we shy away from general statements about the likelihood of the formation of a detonation in GCD models in general.
Our particular model represents only one specific realization, albeit with a choice of offset of the initial deflagration bubble that is already most favourable for a detonation.
For different setups or ignition parameters, for example different choices for the
central density or composition, the situation may be different. 

The progression of the explosion after the onset of the detonation is
visualized in Fig.~\ref{fig:gcd3}. The initialization of the detonation, which occurs 
at $t \approx \unit[2.37]{s}$, is shown in the upper left panel of
Fig.~\ref{fig:gcd3}, where the detonation region is encircled. We see
that the detonation front propagates toward the central region of the
white dwarf, which is predominantly composed of unburned
\nuc{16}{O} and \nuc{12}{C}. Since the central density is still rather
high, most of the stellar material overrun by the detonation is burned
to iron group elements (IGEs), in particular to \nuc{56}{Ni}.
After burning ceases, the detonation ashes are surrounded by the
deflagration ashes for most lines of sight, except for a small opening
angle of reduced contamination around the detonation initiation site; see Fig.~\ref{fig:gcd3}(f).

In Fig.~\ref{fig:energies}, we show the temporal evolution of the
total ($E_\mathrm{tot}$),
nuclear ($E_\mathrm{nuc}$), kinetic ($E_\mathrm{kin}$), internal ($E_\mathrm{int}$) and gravitational energy ($E_\mathrm{grav}$)
as functions of
time. Energy conservation demands $E_\mathrm{tot} = E_\mathrm{kin} +
E_\mathrm{int} + E_\mathrm{egrav} - E_\mathrm{nuc}$. Note that the sub-grid scale
energy, which is of the order of $10^{47}$ to $10^{48}\,\mathrm{erg}$
in turbulent deflagrations, is neglected here \citep[see Figure 1
in][]{schmidt2006c}. Fig.~\ref{fig:energies}(a) shows model GCD200,
whereas Fig.~\ref{fig:energies}(b) shows the energies for the
corresponding pure deflagration model, where no detonation is
initialized. For the GCD simulation, the nuclear energy release is
sufficient to unbind the entire white dwarf
($E_\mathrm{grav}$ asymptotically approaches zero in the late
explosion phase). If no detonation occurs, we obtain a supernova
explosion that fails to fully unbind the WD and leaves behind a
compact remnant after the deflagration flame extinguishes
\citep{jordan2012b,fink2014a}. Such events have been shown to
reproduce the observable characteristics of SN 2002cx-like SNe
\citep{kromer2013a}.

\section{Nucleosynthesis}
Our postprocessing calculations are based on the 384 isotopes nuclear
reaction network YANN \citep{pakmor2012b} that is run on the thermodynamic trajectories
recorded by the tracer particles. The network employs the reaction
rates from the JINA database \citep{cyburt2010a} as of January 27,
2014. As outlined in \citet{kromer2013a}, we implement the effects
of ``solar metallicity'' of the progenitor by initializing the tracer
particle composition to $50\%$\nuc{16}{O}, $48.29\%$\nuc{12}{C}, and
the remaining 1.71\% of the mass to the solar composition from
\citet{asplund2009a} for elements heavier than He, with the exception
of solar C, N, O, which we convert to \nuc{22}{Ne} to account for the
effects of core He-burning. 

\begin{table}
  \centering
  \caption{Stable isotopes (asymptotic values) of model GCD200 in solar masses.}
  \label{tab:syields}
  \begin{tabular}{lclc}
    \hline
    Isotope & Mass      & \hspace{12pt} Isotope & Mass  \\
            & $[\msun]$ &                       & $[\msun]$  \\
    \hline
    \nuc{1}{H}    & 6.19E$-14$   & \hspace{12pt} \nuc{2}{H}   &  0.00E$+00$ \\
    \nuc{3}{He}   & 0.00E$+00$   & \hspace{12pt} \nuc{4}{He}  &  1.29E$-03$ \\
    \nuc{6}{Li}   & 5.55E$-21$   & \hspace{12pt} \nuc{7}{Li}  &  1.61E$-19$ \\
    \nuc{9}{Be}   & 0.00E$+00$   & \hspace{12pt} \nuc{10}{B}  &  7.92E$-16$ \\
    \nuc{11}{B}   & 4.40E$-12$   & \hspace{12pt} \nuc{12}{C}  &  1.88E$-02$ \\
    \nuc{13}{C}   & 6.62E$-08$   & \hspace{12pt} \nuc{14}{N}  &  1.49E$-05$ \\
    \nuc{15}{N}   & 2.04E$-08$   & \hspace{12pt} \nuc{16}{O}  &  1.13E$-01$ \\
    \nuc{17}{O}   & 1.98E$-06$   & \hspace{12pt} \nuc{18}{O}  &  2.54E$-08$ \\
    \nuc{19}{F}   & 2.45E$-10$   & \hspace{12pt} \nuc{20}{Ne} &  6.83E$-03$ \\
    \nuc{21}{Ne}  & 1.64E$-06$   & \hspace{12pt} \nuc{22}{Ne} &  6.73E$-04$ \\
    \nuc{23}{Na}  & 1.16E$-04$   & \hspace{12pt} \nuc{24}{Mg} &  1.96E$-02$ \\
    \nuc{25}{Mg}  & 1.68E$-04$   & \hspace{12pt} \nuc{26}{Mg} &  2.46E$-04$ \\
    \nuc{27}{Al}  & 1.00E$-03$   & \hspace{12pt} \nuc{28}{Si} &  2.52E$-01$ \\
    \nuc{29}{Si}  & 1.08E$-03$   & \hspace{12pt} \nuc{30}{Si} &  2.71E$-03$ \\
    \nuc{31}{P}   & 5.39E$-04$   & \hspace{12pt} \nuc{32}{S}  &  1.02E$-01$ \\
    \nuc{33}{S}   & 2.36E$-04$   & \hspace{12pt} \nuc{34}{S}  &  2.82E$-03$ \\
    \nuc{36}{S}   & 2.63E$-07$   & \hspace{12pt} \nuc{35}{Cl} &  1.29E$-04$ \\
    \nuc{37}{Cl}  & 2.15E$-05$   & \hspace{12pt} \nuc{36}{Ar} &  1.69E$-02$ \\
    \nuc{38}{Ar}  & 1.21E$-03$   & \hspace{12pt} \nuc{40}{Ar} &  8.37E$-09$ \\
    \nuc{39}{K}   & 6.58E$-05$   & \hspace{12pt} \nuc{41}{K}  &  3.58E$-06$ \\
    \nuc{40}{Ca}  & 1.49E$-02$   & \hspace{12pt} \nuc{42}{Ca} &  2.45E$-05$ \\
    \nuc{43}{Ca}  & 2.91E$-08$   & \hspace{12pt} \nuc{44}{Ca} &  9.28E$-06$ \\
    \nuc{46}{Ca}  & 2.58E$-11$   & \hspace{12pt} \nuc{48}{Ca} &  9.40E$-16$ \\
    \nuc{45}{Sc}  & 1.34E$-07$   & \hspace{12pt} \nuc{46}{Ti} &  1.32E$-05$ \\
    \nuc{47}{Ti}  & 4.58E$-07$   & \hspace{12pt} \nuc{48}{Ti} &  3.09E$-04$ \\
    \nuc{49}{Ti}  & 2.31E$-05$   & \hspace{12pt} \nuc{50}{Ti} &  1.91E$-10$ \\
    \nuc{50}{V}   & 1.30E$-09$   & \hspace{12pt} \nuc{51}{V}  &  7.91E$-05$ \\
    \nuc{50}{Cr}  & 3.31E$-04$   & \hspace{12pt} \nuc{52}{Cr} &  7.29E$-03$ \\
    \nuc{53}{Cr}  & 7.40E$-04$   & \hspace{12pt} \nuc{54}{Cr} &  1.25E$-07$ \\
    \nuc{55}{Mn}  & 4.91E$-03$   & \hspace{12pt} \nuc{54}{Fe} &  2.96E$-02$ \\
    \nuc{56}{Fe}  & 7.43E$-01$   & \hspace{12pt} \nuc{57}{Fe} &  1.84E$-02$ \\
    \nuc{58}{Fe}  & 7.04E$-08$   & \hspace{12pt} \nuc{59}{Co} &  3.45E$-04$ \\
    \nuc{58}{Ni}  & 3.74E$-02$   & \hspace{12pt} \nuc{60}{Ni} &  1.50E$-03$ \\
    \nuc{61}{Ni}  & 6.47E$-05$   & \hspace{12pt} \nuc{62}{Ni} &  5.24E$-04$ \\
    \nuc{64}{Ni}  & 1.10E$-13$   & \hspace{12pt} \nuc{63}{Cu} &  3.08E$-07$ \\
    \nuc{65}{Cu}  & 5.85E$-08$   & \hspace{12pt} \nuc{64}{Zn} &  9.31E$-07$ \\
    \nuc{66}{Zn}  & 1.44E$-06$   & \hspace{12pt} \nuc{67}{Zn} &  7.72E$-10$ \\
    \nuc{68}{Zn}  & 4.70E$-10$   & \hspace{12pt} \nuc{70}{Zn} &  4.81E$-24$ \\
    \nuc{69}{Ga}  & 4.98E$-15$   & \hspace{12pt} \nuc{71}{Ga} &  7.24E$-16$ \\
    \nuc{70}{Ge}  & 1.50E$-15$   & \hspace{12pt}              &             \\
    \hline
  \end{tabular}
\end{table}

\begin{table}
  \centering
  \caption{Radioactive isotopes of model GCD200 at $t=100\,\mathrm{s}$
    in solar
    masses.}
  \label{tab:ryields}
  \begin{tabular}{lclc}
    \hline
    Isotope & Mass      & \hspace{12pt} Isotope & Mass  \\
            & $[\msun]$ &                       & $[\msun]$  \\
    \hline
     \nuc{14}{C} & 1.14E$-05$ & \hspace{12pt} \nuc{22}{Na} & 9.39E$-09$ \\
    \nuc{26}{Al} & 1.29E$-06$ & \hspace{12pt} \nuc{32}{Si} & 2.18E$-08$ \\
     \nuc{32}{P} & 5.36E$-07$ & \hspace{12pt}  \nuc{33}{P} & 3.98E$-07$ \\
     \nuc{35}{S} & 5.25E$-07$ & \hspace{12pt} \nuc{36}{Cl} & 6.50E$-07$ \\
    \nuc{37}{Ar} & 2.08E$-05$ & \hspace{12pt} \nuc{39}{Ar} & 7.73E$-09$ \\
     \nuc{40}{K} & 3.86E$-08$ & \hspace{12pt} \nuc{41}{Ca} & 3.57E$-06$ \\
    \nuc{44}{Ti} & 9.26E$-06$ & \hspace{12pt}  \nuc{48}{V} & 5.55E$-08$ \\
     \nuc{49}{V} & 2.14E$-07$ & \hspace{12pt} \nuc{48}{Cr} & 3.09E$-04$ \\
    \nuc{49}{Cr} & 2.29E$-05$ & \hspace{12pt} \nuc{51}{Cr} & 3.65E$-06$ \\
    \nuc{51}{Mn} & 7.55E$-05$ & \hspace{12pt} \nuc{52}{Mn} & 3.42E$-06$ \\
    \nuc{53}{Mn} & 4.82E$-05$ & \hspace{12pt} \nuc{54}{Mn} & 1.25E$-07$ \\
    \nuc{52}{Fe} & 7.23E$-03$ & \hspace{12pt} \nuc{53}{Fe} & 6.92E$-04$ \\
    \nuc{55}{Fe} & 1.35E$-04$ & \hspace{12pt} \nuc{59}{Fe} & 1.91E$-15$ \\
    \nuc{60}{Fe} & 7.93E$-18$ & \hspace{12pt} \nuc{55}{Co} & 4.77E$-03$ \\
    \nuc{56}{Co} & 2.40E$-05$ & \hspace{12pt} \nuc{57}{Co} & 2.77E$-05$ \\
    \nuc{58}{Co} & 6.64E$-08$ & \hspace{12pt} \nuc{60}{Co} & 9.99E$-13$ \\
    \nuc{56}{Ni} & 7.42E$-01$ & \hspace{12pt} \nuc{57}{Ni} & 1.84E$-02$ \\
    \nuc{59}{Ni} & 6.93E$-05$ & \hspace{12pt} \nuc{63}{Ni} & 7.71E$-14$ \\
    \nuc{62}{Zn} & 5.24E$-04$ & \hspace{12pt} \nuc{65}{Zn} & 3.53E$-10$ \\
    \nuc{65}{Ga} & 3.97E$-08$ & \hspace{12pt} \nuc{68}{Ge} & 4.70E$-10$ \\
    \hline
  \end{tabular}
\end{table}

\begin{figure*}
  \includegraphics[width=\linewidth]{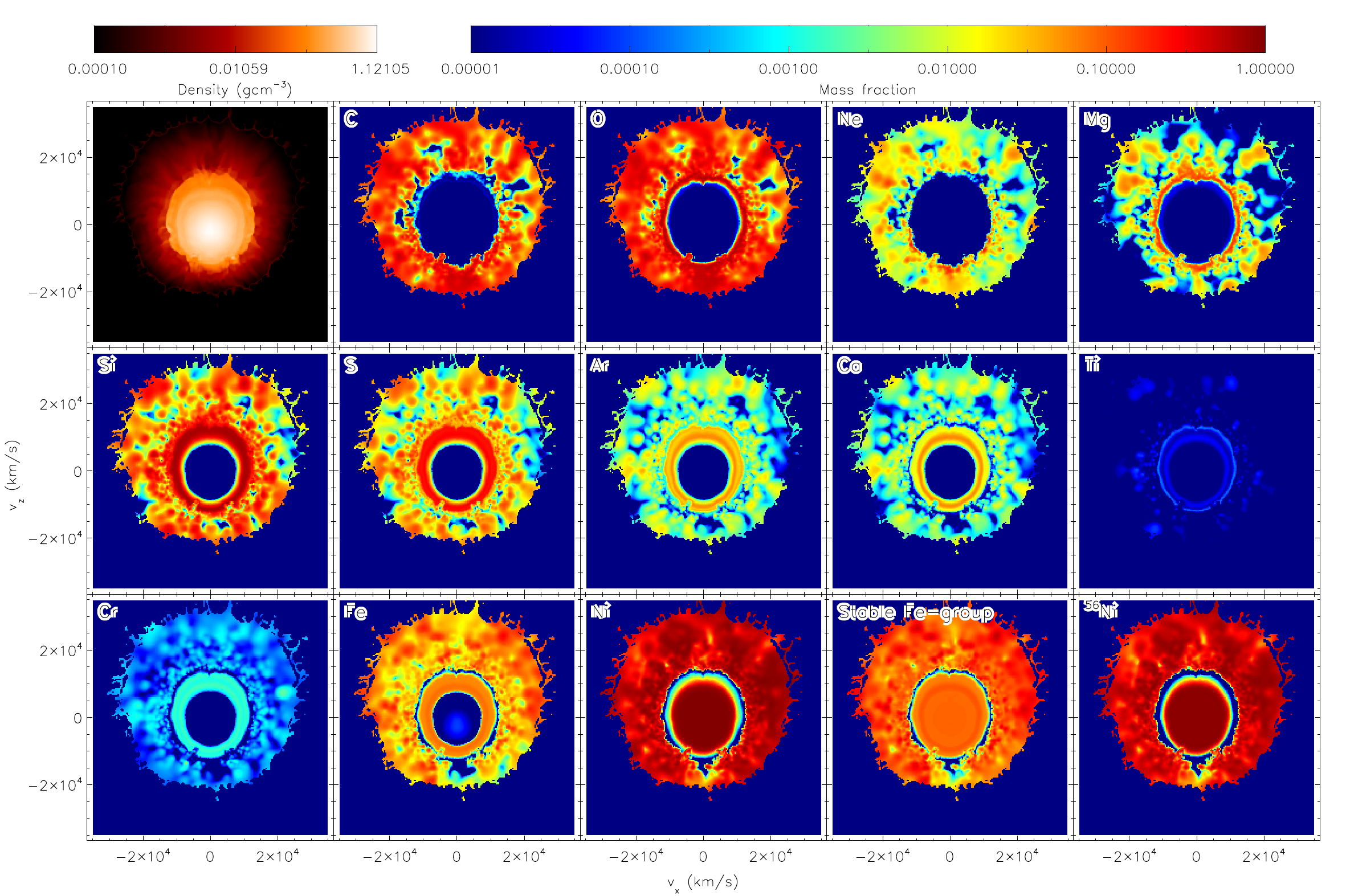}
  \caption{Final ejecta structure of model GCD200 in asymptotic
    velocity space at 100\,s. We show the density (top left panel) and
    the mass fractions of select species in a slice through the
    midplane of the simulation volume along the $x$ and $z$-axes
    (detonation ignition occurred close to the negative $z$-axis).}
  \label{fig:ejecta_composition}
\end{figure*}

\begin{figure*}
  \includegraphics[width=\linewidth, trim= 0 50 0 0, clip]{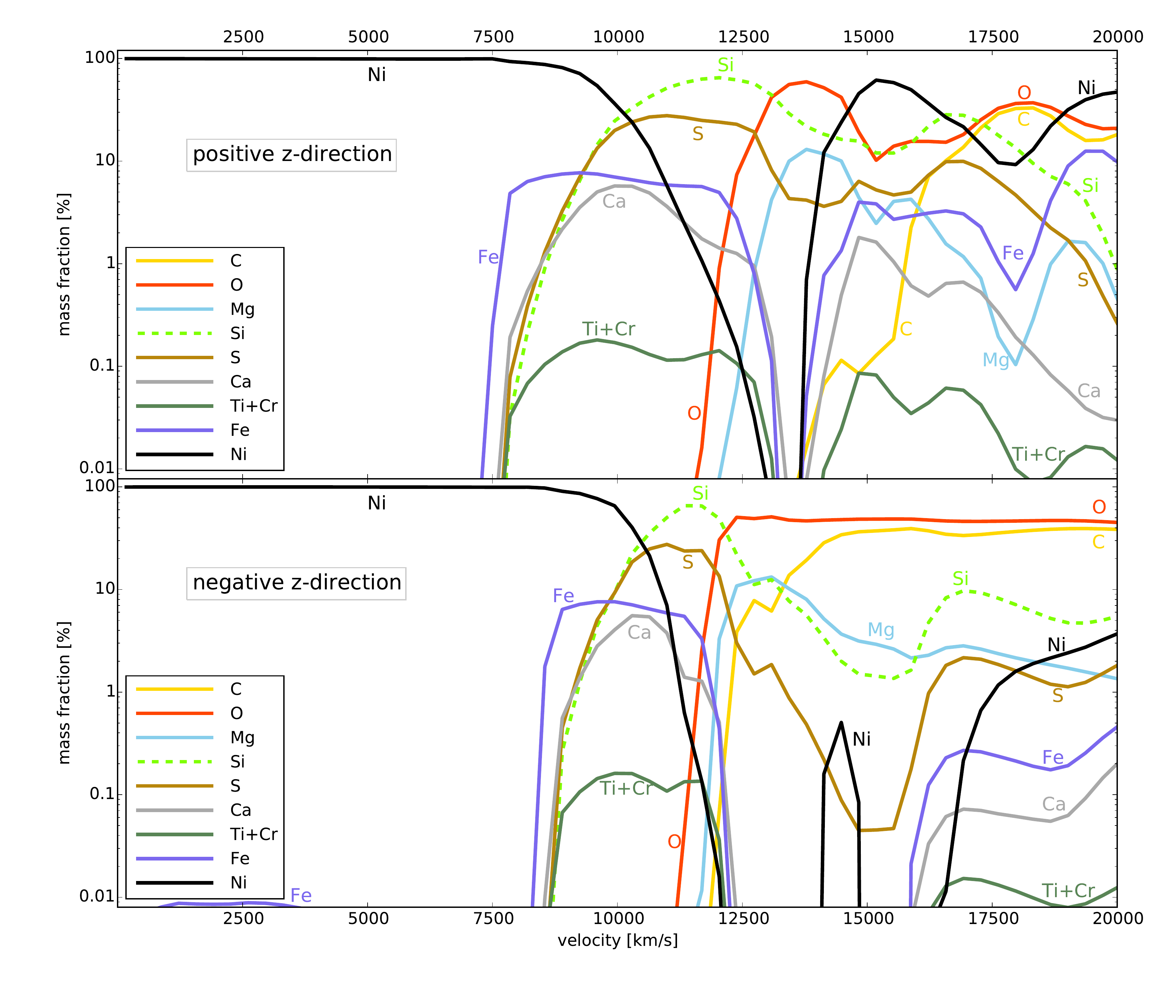}
    \caption{Final ejecta structure of model GCD200 along the positive (top) and negative (bottom) $z$-axis in asymptotic velocity space at 100\,s. Detonation ignition occurred close to the negative $z$-axis. The relative absence of high-velocity IGE in this direction is a reflection of the lack of deflagration ashes (see Fig.~\ref{fig:ejecta_composition}). The profiles along the $\pm x$-axis and $\pm y$-axis qualitatively resemble the top panel.}
  \label{fig:zvelprofile}
\end{figure*}

The masses of stable isotopes after decaying all radioactive nuclides 
are given in Table~\ref{tab:syields}. Radioactive species
$100\,\mathrm{s}$ after ignition are listed in
Table~\ref{tab:ryields}. Burning in both deflagration and detonation
yields $\unit{0.742}{\,\msun}$ of \nuc{56}{Ni} and $\unit{0.030}{\,\msun}$ and
$\unit{0.037}{\,\msun}$ of the stable iron group isotopes \nuc{54}{Fe} and \nuc{58}{Ni}, respectively.
The former is in large parts synthesized in the detonation phase whereas the latter two also formed copiously in the deflagration near the center where high
densities favor neutronization by electron capture
reactions. However, the stable iron group isotopes synthesised in the deflagration do not remain at low velocity. They are carried towards the surface of the WD in the rising deflagration plumes and end up at the highest velocities, see Figs.~\ref{fig:ejecta_composition}~and~\ref{fig:zvelprofile}.

Noteworthy, the GCD200 model has a sub-solar Mn to Fe production ratio of [Mn/Fe]=-0.13. This sub-solar ratio is a reflection of the fact that when the detonation is initiated, the central density of the WD has already fallen below the separation density between ``normal'' and ``$\alpha$-rich'' freeze-out, $\rho < 2 \times 10^8 \gcc$ \citep{thielemann1986a, bravo2012a}, where proton captures during the $\alpha$-rich freeze-out drive the Mn to Fe production ratio lower \citep[e.g.,][]{jordan2003a,seitenzahl2013b}.
Irrespective of occurrence rate and in spite of originating from exploding near-\mch\ WDs, the GCD model of near-\mch\ SN~Ia explosions therefore cannot explain the solar Mn to Fe ratio \citep[see][]{seitenzahl2013b}.

Using a smoothed-particle-hydrodynamics-like algorithm that
approximately conserves the integrated yields \citep[for details,
see][]{kromer2010a}, we map the abundance and density structure of the
SN ejecta at the end ($t=100$\,s) of the hydrodynamic simulations (by
which point homologous expansion is a good approximation, e.g., \citealp{roepke2005c}) to a $200^3$
Cartesian grid.  The resulting ejecta structure is shown in Fig.~\ref{fig:ejecta_composition}. 
We note the asymmetry of the model, with the ejecta in the upper hemisphere extending to much higher velocities. This asymmetric extent of the ejecta in velocity space is a natural consequence of the single spot near-surface lit detonation \citep{chamulak2012a}. The direction of travel of the detonation is pointing into the star on one hemisphere (with a component against the direction of expansion; along gravity) and pointing out of the star on the other hemisphere (with a component along the direction of expansion; against gravity), which leads to the systematic asymmetry of the ejecta.

The total mass of \nuc{56}{Ni} produced is $\unit[0.74]{\msun}$, hence we obtain a bright
explosion. This \nuc{56}{Ni} mass is very close to the $\unit[0.78]{\msun}$ of \nuc{56}{Ni} determined for SN~1991T by \citet{sasdelli2014a}. 
Using the method of ``abundance tomography'' \citep{stehle2005a}, \citet{sasdelli2014a} also find that
they require three per cent of \nuc{56}{Ni} by mass at velocities
above $12,500\,\kms$ to match the spectral time evolution of
SN~1991T.\footnote{Note that the earlier work by \citet{mazzali1995a} found that the outer layers of SN~1991T are dominated by IGEs, which would argue in favor of a GCD origin. However, according to
  \citet{sasdelli2014a}, the large amount of high-velocity \nuc{56}{Ni}
  found by \citet{mazzali1995a} is a result of a larger assumed distance modulus ($\mu=30.65$); \citet{sasdelli2014a} prefer $\mu=30.57$, which leads to improved spectral fits at all photospheric epochs.} To facilitate the comparison with their abundance tomography results (Figures 5 and 7 of \citealt{sasdelli2014a}), we show the chemical ejecta structure in velocity space along two lines of sight ($\pm z$-axes, see Fig.~\ref{fig:zvelprofile}).

The high velocity IGEs pose a problem to our model (when compared to SN~1991T): the shell of deflagration ashes has an IGE content of ${\sim}40$ per cent by mass in almost every direction, which is in strong conflict with \citet{sasdelli2014a}. Furthermore, again in contrast to the \citet{sasdelli2014a} result, the GCD model does not predict a low-velocity core dominated by stable Fe. As expected, neutron-rich, stable Fe-isotopes are produced at the highest densities where electron captures lower the electron fraction. For the GCD model, however, these products of the high-density burning unavoidably rise towards the surface due to buoyancy, even for the case of central igntion \citep{malone2014a}. A consequence of the buoyancy of the hot deflagration ashes is that the stable IGEs end up predominantly at the highest velocities (see Fig.~\ref{fig:ejecta_composition} and top panel of Fig.~\ref{fig:zvelprofile}). In contrast, the low-velocity core is dominated by \nuc{56}{Ni}, which is synthesized in the detonation after the WD has expanded to lower central density. The high velocity deflagration ashes are therefore a characteristic feature of the underlying explosion mechanism of the model, which indeed relies on the convergent flow of the deflagration products on the surface of the WD to trigger the detonation. The distribution of the intermediate mass elements on the other hand generally agrees quite well for a few prominent species such as O, S, Si.

There is, however, more intriguing agreement between the special line of sight along the negative $z$-axis (bottom panel of Fig.~\ref{fig:zvelprofile}) and several features of the abundance tomography results from \citet{sasdelli2014a}. Since the deflagration ashes failed to fully surround the region where the detonation initiated, this particular line of sight is less contaminated by high-velocity deflagration ashes, which improves the comparison. Si and S for example peak between $10,000\,\kms$ and $13,000\,\kms$, O dominates the outer ejecta at high velocity ($>12,500\,\kms$), and \nuc{56}{Ni} is present at high velocity at the few per cent level, all in good agreement with the tomography results of \citet{sasdelli2014a}. The discrepancy concerning the absence of appreciable low-velocity stable Fe however remains, also C is much more abundant at high velocity in the model compared to the tomography results. Still, this is the line of sight where the model resembles the observationally inferred abundance stratification the most.  However, there is only a small solid angle of ${\sim}0.38\,\mathrm{sr}$ (corresponding to a cone of half-opening angle of ${\lesssim}20^{\circ}$) where the deflagration ashes are nearly absent from the surface (see Fig.~\ref{fig:ejecta_composition}). This strongly argues against an identification of this special viewing direction with SNe~91T: a view of the supernova from a line of sight that intersects the deflagration ashes is more than thirty times more likely, and those viewing angles are lacking potential SN counterparts.
In the next section, we present the results of our radiative transfer calculation for the GCD200 explosion model and compare the synthetic observables to SN~1991T. 

\section{Synthetic observables}
\label{sec:obs}
To obtain synthetic spectra and light curves for the GCD200 model,
we use the time-dependent 3D Monte Carlo radiative transfer code
\textsc{artis} \citep{kromer2009a,sim2007b}.  For our radiative
transfer simulation, we remap the ejecta structure to a $50^3$ grid and
follow the propagation of $10^8$ photon packets for 111
logarithmically-spaced time steps between 2 and 120\,d after
explosion.  To reduce the computational costs, a grey approximation,
as discussed by \citet{kromer2009a}, is used in optically thick cells,
and for $t < 3$~d local thermodynamic equilibrium is assumed.  We use
the atomic data set described in \citet{gall2012a}.
Synthetic spectra for 121 equally-sized viewing-angle bins covering
the full solid angle are extracted on a 1000-bin logarithmic
wavelength grid spanning a range between 600 and 30,000 \AA.  

\begin{figure}
  \includegraphics[width=\linewidth]{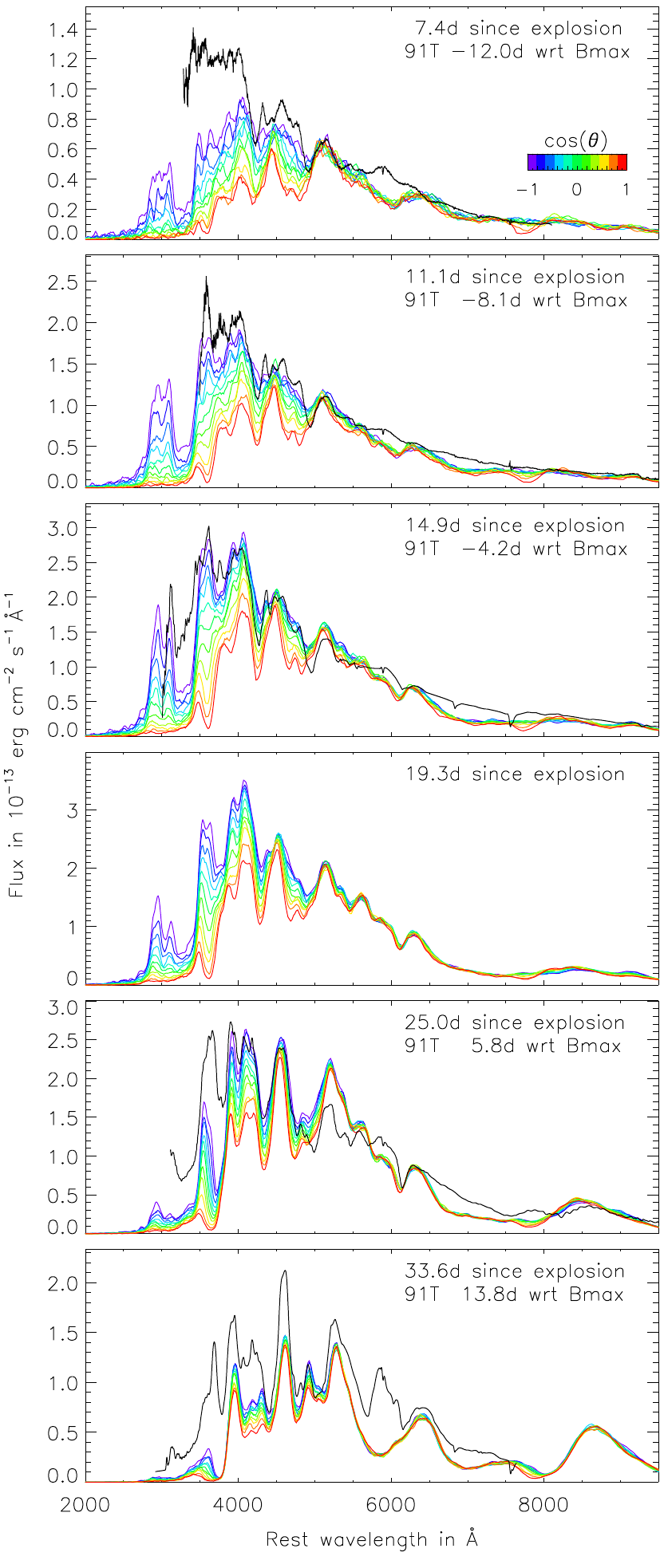}
  \caption{Time sequence of synthetic spectra for our model (lines of
    different colours correspond to different viewing angles as
    indicated by the colour bar). The snapshot at 19.3\,d corresponds
    roughly to $B$-band maximum in the model, which occurs between
    19.2 and 19.9\,d depending on the viewing angle. The synthetic
    spectra have been smoothed with a Savitzky-Golay filter to reduce
    Monte Carlo noise. For comparison we overplot observed spectra of
    SN~1991T at corresponding epochs (black, the flux calibration has
    been checked against the photometry and adjusted if
    necessary). The observed spectra have been de-reddened
    ($E(B-V)=0.13$, \citealt{phillips1992a}) and de-redshifted
    ($z=0.006059$, from interstellar Na). We assumed a distance
    modulus of 30.76 to SN~1991T \citep{saha2006a}.}
  \label{fig:specseq}
\end{figure}

On large scales, the GCD200 model is roughly symmetric under rotation about the axis
defined by the center of the star and the location of the initial deflagration bubble
(the $z$-axis in Fig.~\ref{fig:ejecta_composition}).
Thus, we can integrate the synthetic spectra over the equatorial angle with
respect to the $z$-axis, leaving 11 equally-sized viewing-angle bins
in $\cos\theta$. Fig.~\ref{fig:specseq} shows snapshots of the
spectral evolution of our model from 7.4\,d to 33.6\,d past explosion.

At early times ($t\lesssim25$\,d), the model spectra show a prominent
viewing-angle sensitivity in the ultraviolet (UV) and blue wavelength
regions where the flux level decreases from $\cos\theta=-1$ to
$\cos\theta=1$ by up to a factor 10 at certain wavelengths. This is
due to the asymmetric distribution of the IGE-rich deflagration ashes:
on the deflagration ignition side ($\cos\theta=1$) IGEs are abundant
up to ${\sim}30,000\kms$, whereas the ejecta extend only to ${\sim}20,000\kms$ on the far side ($\cos\theta=-1$) where
the detonation ignites (see Fig.~\ref{fig:ejecta_composition}). This
leads to a significant reduction of IGE line blanketing from
$\cos\theta=1$ to $\cos\theta=-1$ and thus an enhancement of the flux
in the UV and blue regions when viewed from the detonation side.
Redder wavelengths ($\lambda \gtrsim 5000$\,\AA) are not affected by
line blanketing from IGEs and thus show no prominent viewing-angle
sensitivity. At later times ($t\gtrsim 25$\,d), the outer layers become
optically thin and the viewing-angle sensitivity in the blue bands
decreases. The remaining viewing-angle sensitivity is mainly due to
the off-centre structure of the detonation ashes (see
Fig.~\ref{fig:ejecta_composition}).

\begin{figure}
  \includegraphics[width=\linewidth]{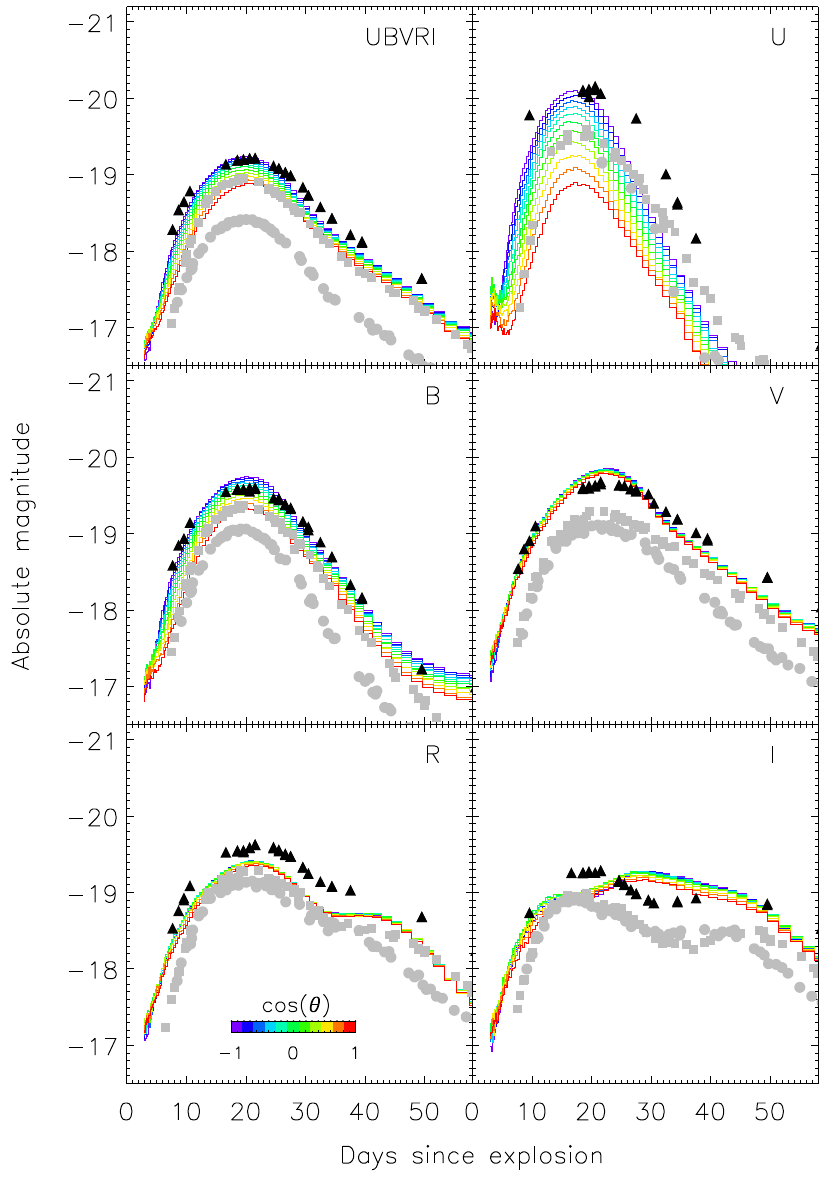}
  \caption{Synthetic broadband light curves for our model. Lines of
    different colour correspond to different viewing angles as
    indicated by the colour bar. Overplotted are observed light curves
    for the normal SNe~Ia 2004eo (grey circles) and 2005cf (grey
    squares), and for the overluminous SN 1991T (black triangles). For
    the comparison we assume a $B$-band rise time of 19.6\,d, which
    corresponds to the average $B$-band rise time of our model.}
  \label{fig:lcs}
\end{figure}

The same behaviour can also be observed in the synthetic light curves
(Fig.~\ref{fig:lcs}), which show a large variation around peak in the
$U$ and $B$ bands. For $U$, the peak magnitudes vary between -18.9 and
-20.1\,mag with rise times between 16.8 and 17.5\,d. The $B$-band
light curves peak between 19.2 and 19.9\,d at magnitudes between -19.3
and -19.8\,mag. In contrast, we do not observe any significant
variation in $V$, $R$ and $I$. After maximum the viewing angle
sensitivity also decreases in the bluer bands.

The $U$- and $B$-band light curves show an additional effect at very
early epochs ($t\lesssim5$\,d). At these epochs $\gamma$-rays from
\nuc{56}{Ni} decay in the deflagration ashes are still trapped,
leading to additional surface heating and an enhanced flux in the
UV. This causes a decline in the $U$-band light curves at the earliest
epochs and a plateau-like behaviour in the $B$ band.

Fig.~\ref{fig:lcs} also shows light curves for three observed
SNe~Ia. The luminous SN~1991T \citep{lira1998a} as well as SN~2004eo \citep{pastorello2007b} and SN~2005cf
\citep{pastorello2007a}, which are representative for the class of normal
SNe~Ia. Our model light curves are not a particularly
good fit to any of these objects. Compared to normal SNe~Ia, the model
light curves are clearly too bright. This is no surprise given that
the \nuc{56}{Ni} mass of our model (0.74\,\msun) is significantly
larger than the typical range of 0.4 to 0.6\,\msun\ inferred for most SNe~Ia
\citep{scalzo2014a}.  Compared to SN~1991T, an equatorial viewing
angle ($\cos\theta=0$) is in reasonable agreement with the observed
$B$- and $V$-band light curves for $t\lesssim10$\,d and
$t\gtrsim30$\,d. However, around peak a viewing angle close to the
deflagration ignition side agrees better with the observed $B$-band
light curve of SN~1991T, while in the $V$-band our model is
significantly too bright for all viewing angles.

Looking at our model spectra, we also find no convincing match
with observed SN~Ia spectra. Compared to normal SNe~Ia the line
features associated with intermediate-mass elements (IMEs) are
slightly too weak. \ions{Si}{ii} $\lambda5972$, for example, is very
weak in the model spectra indicating relatively high ionization and
temperature of the ejecta. This is corroborated by the presence of
absorption features at $\sim 4200$ and 4900\,\AA\ that are attributed
to \ions{Fe}{iii}. Compared to SN~1991T there is no individual viewing
angle that provides good agreement over the full wavelength range and
multiple epochs (see Fig.~\ref{fig:specseq}). Although relatively weak, the IME
features of our model are still too strong at pre-maximum epochs when
SN~1991T shows no clear signs of the \ions{Ca}{ii} H\&K lines and
\ions{S}{ii} $\lambda5624$.  \ions{Si}{ii} $\lambda6355$ develops in
SN~1991T a week before maximum, but it is also weaker than in our
model.

\section{Summary and Conclusions}
\label{sec:summary}
The purpose of this work is to calculate what the generic observables of a GCD model look like \emph{if} a detonation occurs,
not to settle the question \text{whether} a detonation can form in the first place. Deflagrations that are ignited closer to the centre lead to more burning and expansion, which leads to a weaker collision of the deflagration ashes at the opposite point of break-out. To obtain detonation conditions and a subsequent GCD SN explosion, we ignited a near-\mch\ WD outside the preferred range predicted by numerical simulations \citep[][]{nonaka2012a} and chose less stringent critical detonation conditions compared to previous works \citep[cf.][]{roepke2007a}.

Our high resolution model ($512^3$) was ignited (200\,km) and
fullfilled our deliberately optimistic critical detonation conditions and we
initialized a detonation and followed the explosion until
homologous expansion. Given the large offset, relatively little mass burned in the deflagration, which results in a relatively compact WD at the instant of detonation initiation. This means that GCD explosions are automatically linked to the brighter end of the SN~Ia distribution.
If the WD were ignited closer to the center, we would get a stronger deflagration, resulting in more expansion, no detonation, and a sub-luminous SN~2002cx-like event \citep{jordan2012b,kromer2013a}.

By post-processing one million tracer 
particles with a nuclear reaction network, we
determined the detailed nucleosynthesis in the explosion.
A high-velocity, outer shell of deflagration ashes rich in \nuc{56}{Ni} and stable iron group isotopes such as \nuc{54}{Fe} and \nuc{58}{Ni} is a generic and robust prediction for this class of models. This shell of IGE-rich deflagration ashes exhibits a gradient in its extent in velocity space, decreasing in extent from the deflagration-ignition side to the far side where the detonation-initiation side.

We used the resulting three-dimensional composition structure of the explosion ejecta
as input for a Monte-Carlo radiative transfer calculation with the \ARTIS\ code
to obtain time-dependent spectra and light curves for the GCD200 model.  
Comparing these synthetic observables to SN~Ia data, we conclude
that this GCD model cannot explain any of the more common sub-classes 
of SNe~Ia. In particular, it falls short
of explaining SN 1991T-like events, a class of SNe that owing to their
brightness and early high-velocity Fe-features appeared to be the most
natural candidate \citep{fisher2015a}. 
First, the single spot ignition of the deflagration on one side of the star and the initiation of the detonation on the opposite side naturally leads to an asymmetric extent of the shell of deflagration ashes in velocity space, extending to much higher velocity on the side where the deflagration was ignited. This in turn leads to a strong viewing angle dependence for the synthetic spectra and light curves, most pronounced in the bluer bands at early times, which is not in agreement with observations. Since the ejecta asymmetry is a generic feature of GCD explosions \citep{chamulak2012a}, the resulting viewing angle sensitivity is generic as well.
Identification of SNe~91T with only the peculiar line of sight towards the location of the initiation of the detonation, which gives the best agreement, leads to the following problem: the other viewing angles characterized by high-velocity, turbulently mixed deflagration ashes rich in \nuc{56}{Ni} and stable IGE statistically dominate and do not correspond to a known, more common sub-class of SNe.
Second, our model is too abundant in IGEs at high velocity and lacks the stable low-velocity Fe inferred by \citet{sasdelli2014a} by abundance tomography methods.
Third, although the overall flux-levels compare favourably, the spectral features do not. For example the \ions{Ca}{ii} H\&K lines, 
\ions{S}{ii} $\lambda5624$, and \ions{Si}{ii} $\lambda6355$, which are absent or very weak pre-maximum in SN~1991T, are too strong in our model for all lines of sight.
Particularly the first two points are rather robust, generic features of GCD explosion models. Overall, our results therefore suggest that GCD is probably not a good explosion model for SN~1991T. Whether or not more peculiar events can be explained by GCD events remains to be seen.

\section*{Acknowledgements}
This work was supported by Australian Research Council Laureate Grant FL0992131, the Deutsche Forschungsgemeinschaft via the Transregional Collaborative Research Center TRR 33 ``The Dark Universe'', the Emmy Noether Program (RO
3676/1-1), the ARCHES prize of the German Ministry of Education and
Research (BMBF), the graduate school ``Theoretical Astrophysics and
Particle Physics'' at the University of W\"urzburg (GRK 1147) and the
Excellence Cluster EXC~153 ``Origin and Structure of the Universe''. SAS acknowledges support from STFC grant ST/L000709/1. AJR is thankful for funding provided by the Australian Research Council Centre of Excellence for All-sky Astrophysics (CAASTRO) through project number CE110001020. RP acknowledges support by the European Research Council under ERC-StG grant EXAGAL-308037. We also thank the DAAD/Go8 German-Australian exchange programme for travel support. STO acknowledges support from Studienstiftung des deutschen Volkes. The work of STO, RP, and FKR is supported by the Klaus Tschira Foundation.

This research was supported by the Partner Time Allocation (Australian National University), the National Computational Merit Allocation and the Flagship Allocation Schemes of the NCI National Facility at the Australian National University.
The authors also gratefully acknowledge the Gauss Centre for Supercomputing
(GCS) for providing computing time through the John von Neumann
Institute for Computing (NIC) on the GCS share of the supercomputer
JUQUEEN \citep{stephan2015a} at J\"ulich Supercomputing Centre (JSC). GCS is the alliance
of the three national supercomputing centres HLRS (Universit\"at
Stuttgart), JSC (Forschungszentrum J\"ulich), and LRZ (Bayerische
Akademie der Wissenschaften), funded by the German Federal Ministry of
Education and Research (BMBF) and the German State Ministries for
Research of Baden-W\"urttemberg (MWK), Bayern (StMWFK), and
Nordrhein-Westfalen (MIWF). 

\bibliography{astrofritz}
\bibliographystyle{aa}

\end{document}